# A COVID-19 Search Engine (CO-SE) with Transformer-based Architecture

Shaina Raza[1*]

[1] University of Toronto, Toronto, ON, Canada.

shaina.raza@utoronto.ca

**Abstract**

Coronavirus disease (COVID-19) is an infectious disease, which is caused by the SARS-CoV-2 virus. Due to the growing literature on COVID-19, it is hard to get precise, up-to-date information about the virus. Practitioners, front-line workers, and researchers require expert-specific methods to stay current on scientific knowledge and research findings. However, there are a lot of research papers being written on the subject, which makes it hard to keep up with the most recent research. This problem motivates us to propose the design of the COVID-19 Search Engine (CO-SE), which is an algorithmic system that finds relevant documents for each query (asked by a user) and answers complex questions by searching a large corpus of publications. The CO-SE has a retriever component trained on the TF-IDF vectorizer that retrieves the relevant documents from the system. It also consists of a reader component that consists of a Transformer-based model, which is used to read the paragraphs and find the answers related to the query from the retrieved documents. The proposed model has outperformed previous models, obtaining an exact match ratio score of 71.45% and a semantic answer similarity score of 78.55%. It also outperforms other benchmark datasets, demonstrating the generalizability of the proposed approach.

**Keywords:** CORD-19; COVID-19; Deep learning; Transformer models; Search engine

## 1. Introduction

Coronavirus disease (COVID-19) is an infectious disease caused by the SARS-CoV-2 virus (Yuki, Fujiogi, and Koutsogiannaki 2020). COVID-19 has affected a lot of people all over the world. It has been reported in about 200 countries and territories, with more than 266 million cases reported around the world and 5.6 million deaths (Google news 2022). COVID-19 has caused many physical complications, such as pneumonia, acute respiratory distress syndrome (ARDS), multi-organ failure, and death. It has also increased mental health issues such as depression, post-traumatic stress disorder, and suicide (Jenkins et al. 2021). A significant number of patients with COVID-19 also report prolonged symptoms, known as long-COVID (McMahon et al. 2021), which can damage the lungs, heart and brain increasing the risk of long-term health problems (Akbarialiabad et al. 2021). As of July 2021, the "long-COVID" symptoms, also known as post-COVID conditions, are classified as a disability under the Americans with Disabilities Act (ADA) (CDC 2021). In this study, we focus on the ongoing pandemic issue and propose an artificial intelligence (AI) solution that can be easily extended and adapted to quickly learn relevant insights critical for understanding and combating any new infectious disease.

In less than two years, there has been an explosion of literature on COVID-19 (since its inception) (Chen, Allot, and Lu 2021; Lu Wang et al. 2020). COVID-19 researchers are facing a significant challenge in sifting through a large body of literature to find relevant and credible information (Raza and Ding 2022). Given the volume of data available on COVID-19, the research community and health care professionals need to have access to timely information as soon as it is available in the literature. However, due to information overload, the physicians and research community frequently struggle to find an instant response to the numerous real-world concerns they encounter. Furthermore, by the time the necessary information is presented, a substantial amount of new research has been published in the literature.

Search engines (Baeza-Yates and Ribeiro-Neto 1999) and the question-answering systems (Bouziane et al. 2015) primarily serve as filters for the vast amount of information available on the internet. These software applications allow users to quickly and easily locate content that is truly relevant or valuable, without having to wade through a plethora of irrelevant documents (research papers, reports). There are some question-answering systems (Alzubi et al. 2021; Ngai et al. 2021; Tang et al. 2020) lately released to filter large amounts of publications data to respond to COVID-19 topics, however, the majority of them focus on pre and/or mid-covid literature. We are taking this research a step further, by also including the literature related to long-term COVID (in addition to pre/mid-COVID-19 issues), as well as investigating the impacts of COVID-19 on public health.

In this work, we propose a COVID-19 Search Engine (CO-SE) that retrieves and ranks articles (publications data) on COVID-19 based on user-specified queries. Our CO-SE system uses the dataset on COVID-19 related research papers and their full text. To begin, we create a retriever component that acts as a filter and rapidly scans the dataset for a set of candidate publications related to a query. The retrieved documents are then passed to the reader component that reads the text of the retrieved documents and finds answers in response to every query. We use a Transformer-based model (Vaswani et al. 2017) inside the reader component. A Transformer (Vaswani et al. 2017) is a deep learning architecture that uses the self-attention mechanism to weigh the significance of each part of the incoming data. The output from the CO-SE is a list of retrieved documents that are ranked, along with answer snippets from those documents. We also provide metadata information with each answer, which includes the document (publication) title, context (surrounding paragraph) of each answer, as well as the offset (starting and ending position) of the answer within the document and in the context. Along with this information, the CO-SE also displays the model's confidence in the accuracy of each returned answer.

By proposing this search engine, we hope to cover a broad range of COVID-19 topics from scholarly literature, including diagnosis, management, vaccines, and long-term COVID, to end the pandemic and prepare for any future pandemics. We list our contributions as:

1. We propose CO-SE, a search engine that allows rapid and easy access to COVID-19-related research publications. To facilitate natural language searches, CO-SE employs machine learning and deep neural network-based techniques. Its functionality includes document ranking and providing contexts and answers in response to any query posed by a user in natural language. Our goal is to make it easier for clinicians, researchers, and other experts to navigate through the massive COVID-19 information and quickly find relevant actionable evidence.

2. We create a COVID-19 publication dataset by curating it through the COVID-19 Open Research Dataset Challenge (CORD-19) (Lu Wang et al. 2020) dataset, sifting through the articles, extracting the text information and creating a database in a structured format (data frames) that includes article text and meta-data information. There are also other COVID-19 publication datasets available in the literature (Chen et al. 2021; Wang and Lo 2021), however, our initial screening of these datasets and a recent review (Wang and Lo 2021) on different COVID-19 datasets revealed to us that there has been an overlap of publications in these repositories, and most of these datasets rely on Allenai[1] COVID-19 scholarly repository as their primary source. As a result, we decided to work with the pioneer source of data i.e., Allenai CORD-19, in this study.

3. We propose to enhance our search engine's results to configure complex search and retrieval of the data through a retriever module. We use a machine learning algorithm Term Frequency-Inverse Document Frequency (TF-IDF) (Aggarwal 2015) in the retriever for the document search and to return candidate documents that are most relevant according to a search query.

4. We go beyond the standard keyword matching to efficiently find relevant answers from the retrieved documents by understanding the query's semantics. To accomplish this, we use a Transformer model (Vaswani et al. 2017) in the reader component. The purpose of implementing a Transformer model in the reader is to derive long-range sequential relationships and a holistic understanding of the text from the publications' data.

5. We prepare a gold-standard dataset on COVID-19 topics. The term 'gold-standard dataset' refers to a set of data that has been prepared and annotated manually by experts (Cardoso et al. 2014). We collaborate with a group of public health experts to manually select and annotate approximately some COVID-19 publications on a variety of topics (epidemiology, long-term COVID, equity, and impacts), resulting in approximately many question-answer pairs in the Stanford Question Answering Dataset (SQuAD) (Rajpurkar et al. 2016) format (a prototypical format for the question-answering task).

To the best of our knowledge, there is no recent gold-standard dataset on COVID-19 in the SQuAD format. The seminal work (Tang et al. 2020; Möller et al. 2020), in this regard, consists of question-answer pairs that are not so recent. Our gold-standard dataset serves two purposes: (1) to fine-tune our Transformer model inside the reader to enhance its reading comprehension capability; and (2) to evaluate the quality of our search engine using a manually annotated dataset. We also fine-tune our Transformer model as well evaluate the performance of the proposed CO-SE reader by testing it on two additional benchmark datasets.

[1] https://allenai.org/data/cord-19

The rest of the paper is organized as follows: section 2 is the related work, section 3 is the data collection section, section 4 is the working of CO-SE Architecture, section 5 is about the experimental setup and section 6 discusses the results and analysis. Section 7 is the discussion, limitation and future directions section and section 8 is the conclusion.

## 2. Related work

### 2.1. COVID-19 datasets

Many COVID-19 datasets are based on the publications' data and represent the scientific information about COVID-19. The COVID-19 Open Research Dataset (CORD-19) (Lu Wang et al. 2020) is one such repository of information. It is a joint challenge launched by the Allen Institute (AI2), the National Institutes of Health (NIH), and the United States federal government through the White House. The CORD-19 challenge is organized by Kaggle and the goal is to extract useful knowledge from thousands of scholarly articles about COVID-19.

LitCovid (Chen, Allot, and Lu 2021), is another open-source dataset that provides centralized access to over 20,000 (and growing) PubMed[2] publications relevant to COVID-19. These datasets are being used for a range of tasks, such as text summarization (Song and Wang 2020), document search (Esteva et al. 2021) and question answering systems (Alzubi et al. 2021; Tang et al. 2020).

COVID-QA (Möller et al. 2020) and CovidQA (Tang et al. 2020) are two small-scale datasets related to COVID-19, which have been annotated by specialists. COVIDRead (Saikh et al. 2021) is another dataset that comprises over 40k manually annotated question-answers. All these COVID QA datasets are usually made available in the SQuAD[3] format. The SQuAD has become a prototypical standard for question-answering systems (C. Yang 2018) and consists of a collection of question-and-answer pairs; where there is a question, and the label is an answer (ground truth label) or the question is unanswerable (impossible as the label). Each answer is also supported by a context (surrounding paragraph).

[2] https://pubmed.ncbi.nlm.nih.gov/
[3] https://rajpurkar.github.io/SQuAD-explorer/

The WHO Global literature on coronavirus disease[4] also provides a portal to search the COVID-19 publications from various scholarly repositories (journals, pre-print servers). However, there is a large overlap of information with the CORD-19 database (Wang and Lo 2021), this is because CORD-19 curates a large portion of the WHO database. There is another curated collection of COVID-19 papers by the Centers for Disease Control and Prevention (CDC)[5]. A large portion of the CDC database overlaps with PubMed and PMC, which are also sources of papers for CORD-19 and LitCovid. The CDC database also contains a collection of white papers and technical reports, which are also found in CORD-19. Other interfaces provide access to COVID-19 papers, such as iSearch[6], Covidex[7], SciSight[8] and others as mentioned in a review article (Chen et al. 2021), these interfaces also rely on the CORD-19 initiative as their primary source of information.

CORD-19 and LitCOVID are the most widely used data sources providing COVID-19 scholarly articles. These two sources also provide researchers with an application programming interface (API) and a file transfer protocol (FTP) server, allowing them to download the data using scripting (mostly bash commands) with inclusion/ exclusion criteria. The researchers mostly use these two data sources for text mining purposes.

### 2.2. COVID-19 models on publications data

The Transformer models are built on transfer learning (Tan et al. 2018) and the neural attention technique (Vaswani et al. 2017). Tang et al. (Tang et al. 2020) have used transfer learning to retrieve the relevant documents from the COVID-19 dataset. Oniani and Wang (Oniani and Wang 2020) have fine-tuned GPT-2 (Wu and Lode 2020) and Möller et al., (Möller et al. 2020) have used the RoBERTa (Liu et al. 2019) on the COVID-19 data for the question answering task. Lee et al. (Lee et al. 2020) proposed a system, COVIDASK, which is a question answering system on COVID-19 that uses text mining tools to build such a system. Wei et al., (Wei et al. 2020) use BERT (Devlin et al. 2018) baseline to classify the questions: transmission, societal effects, preventions and such. Su et al., (Su et al. 2020) propose CAiRE-COVID that pre-processes a user's query and find the most relevant documents. It paraphrases long, complex queries to simple queries that are easier to comprehend by the system. These queries are run through the model, which returns paragraphs from the highest matching score.

[4] https://search.bvsalud.org/global-literature-on-novel-coronavirus-2019-ncov/
[5] https://www.cdc.gov/library/researchguides/2019 novelcoronavirus/researcharticles.html
[6] https://icite.od.nih.gov/covid19/search/
[7] https://covidex.ai/?query=outbreak
[8] https://scisight.apps.allenai.org/

While all these models are highly useful, they are based on early COVID-19 responses. In our work, we use the most relevant and up-to-date collection of COVID-19 publications data. We use a machine learning algorithm (TF-IDF) and a deep neural network-based (Transformer) model to build our search engine. Our goal is to propose a scalable in their search and discovery for answers to high priority scientific queries. Some recent works (Alexandra et al. 2021; McGain et al. 2022; Trent et al. 2022; Chicaiza and Bouayad-Agha 2022; Q. Zhang et al. 2022) on COVID-19 literature and pandemic response are also the motivation for the current study.

## 3. Data Collection

The data collection phase is discussed below.

### 3.1. Dataset from CORD-19

We prepare a COVID-19 dataset by curating an up-to-date collection of COVID-19 articles from CORD-19 (Lu Wang et al. 2020). CORD-19 is a publicly accessible collection of academic articles relating to COVID-19 and related research. The actual CORD-19 dataset was initially made available on March 16, 2020, and is updated daily. It includes articles from PubMed Central, Medline, arXiv, bioRxiv, and medRxiv (Colavizza 2020) as well as from the WHO COVID-19 Database (WHO 2021). We use the latest release[9] of CORD-19 in this work. We apply the following inclusion and exclusion criteria to get data from CORD-19:

- We include only the published literature from CORD-19 in English between 20th March 2020 and 31st December 2021 and exclude papers that are not published in scientific journals (e.g., pre-prints, reports, grey literature). A PMID is a unique integer value assigned to each PubMed[10] record.

[9] https://ftp.ncbi.nlm.nih.gov/pub/lu/LitCovid/litcovid2BioCXML.gz
[10] https://pubmed.ncbi.nlm.nih.gov/

- CORD-19 consists of papers in two formats: Portable Document Format (PDF) and Extensible Markup Language (XML) format. We include only the PDFs of PMC articles that provide the full text, abstract and metadata (title, DOI, etc.) for each article. The XML collection does not provide abstracts of the papers (Colavizza 2020), so we exclude them.
- We exclude the articles in the collection that are the papers before 2020 - CORD-19 also have a portion of the collection in the same timelines (2020-2021) that is before the year 2020 covering historical coronavirus research (Cunningham, Smyth, and Greene 2021)).

After all these filtration criteria, we get around 8,000 CORD-19 unique papers. We parse and convert the PDF, XML, JSON formats of these articles and generate a final output in a Comma-Separated Values (CSV) format with main attributes like ‘PMID’, ‘title’, ‘paragraphs’, ‘URL’, ‘publication date’, ‘authors’. We also specify the complete text of the research articles in the final dataset. The dataset is available in English. The general details of the CORD-19 dataset are given in Table 1:

**Table 1**: General details of CORD-19 used in this work

| Total articles | Articles used | Timeline | Files |
|---|---|---|---|
| Over 50K articles in all formats (PDF, XML) | ~8K (only PMC articles with full text + abstracts) | 2020-03-13 till 2021-12-31 | The dataset consists of the following files [11]: document embeddings for each CORD-19 paper, a collection of JSON files with the full text of CORD-19 papers, and Metadata (title, abstract, text body and other) for all CORD-19 papers. |

While our COVID-19 dataset size may constitute a subset of the actual CORD-19 dataset, this subset is the result of careful filtering. We performed an exploratory analysis of the dataset to determine its structure and discovered that our chosen set of articles covers a wide range of COVID-19-related topics, including epidemiology, public health, health equity, infectious diseases, public policy, and general impacts. Some of the important COVID-19 topics covered in these articles are shown in Figure 1, where we can see, that many articles are related to reviews (scoping, literature, systematic), public health, risk factors, RNA, and health care and so.

[11] https://github.com/allenai/cord19

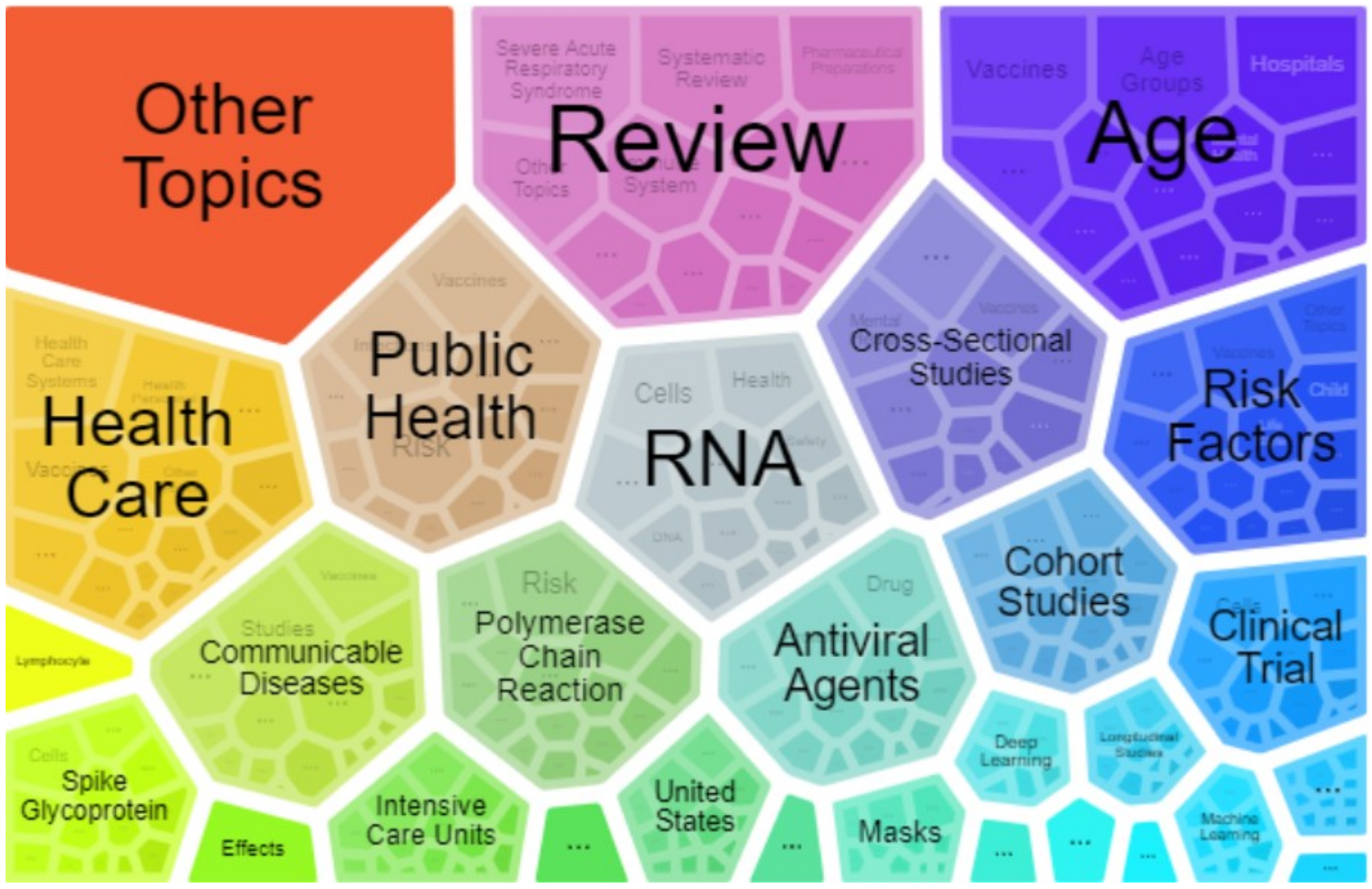

**Figure 1**: Important topics covered in our COVID-19 dataset extracted from CORD-19.

We also group the important topics from our collection of articles as shown in Figure 2. The blue line in Figure 2 shows the coverage of the topic, where we see that epidemiology is the most covered topic, followed by beta coronavirus, asymptomatic, and so on.

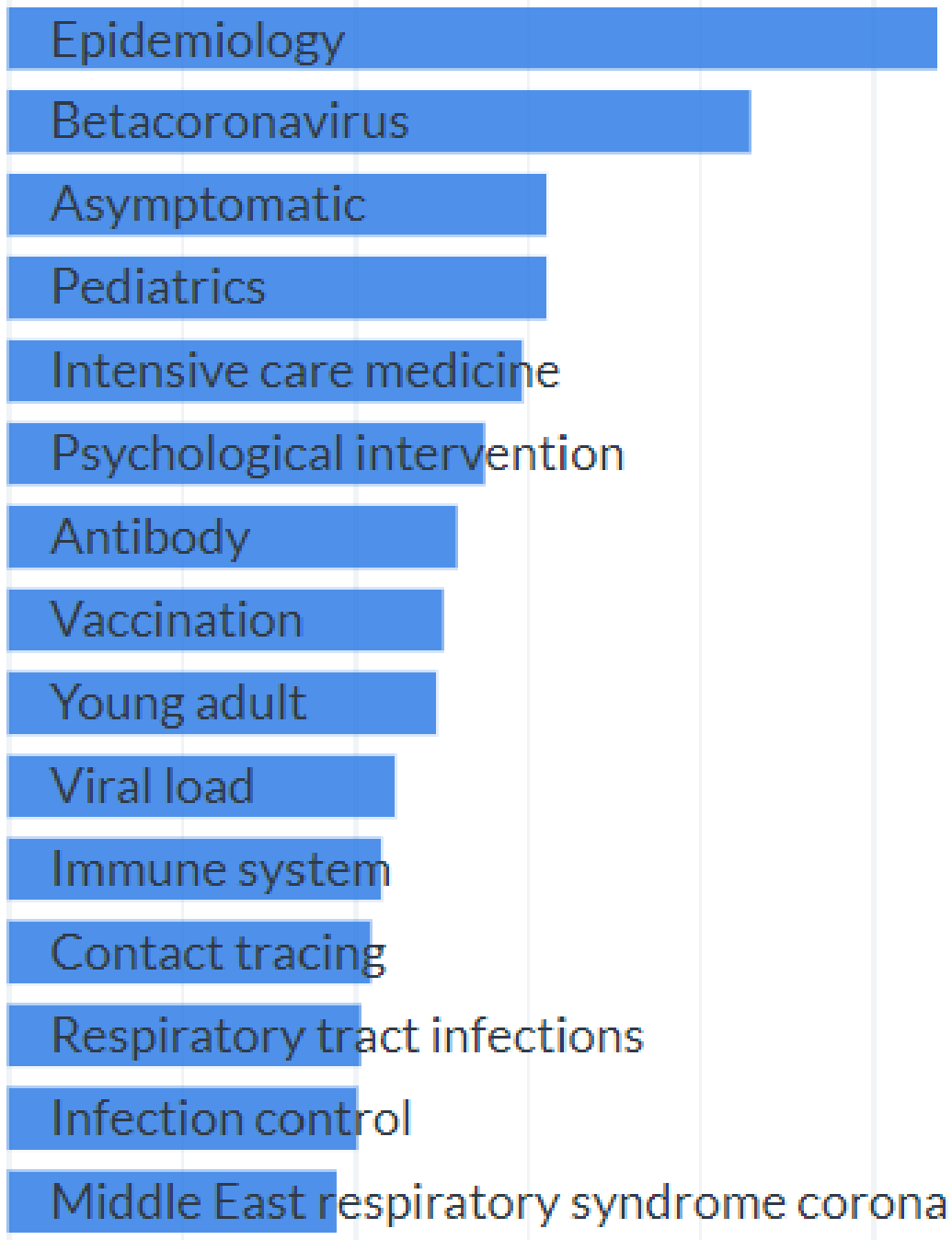

**Figure 2**: Topics covered in the dataset

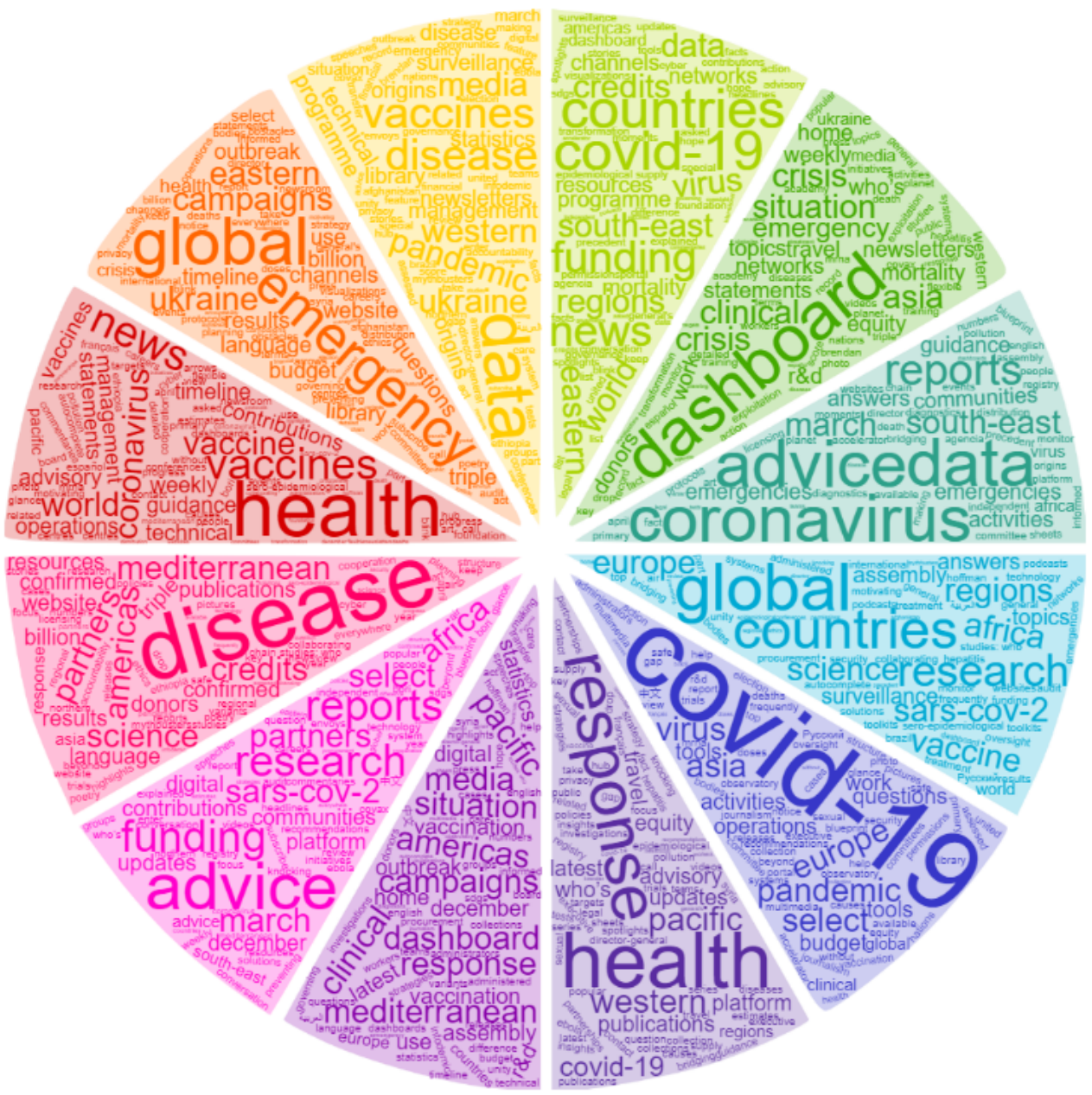

**Figure 3**:  Word cloud from the dataset

We also show the word cloud of important words from our dataset in Figure 3. A word cloud is a visual representation of words where the large word size means the word is covered in more frequency. From this word cloud in Figure 3, we see that some of the words most used in the COVID-19 literature are ‘COVID-19’, ‘health’, ‘disease’, ‘response’ and ‘emergency’.

### 3.2. Other datasets

We also use other datasets in this work for two purposes:

To train the reader module of the CO-SE pipeline (discussed in Section 4); and (2) to evaluate the performance of the CO-SE pipeline.

*A) Our gold-standard dataset CQuAD:*

We prepare a gold-standard dataset that consists of COVID-19 articles manually chosen and annotated by experts in a public health domain. The experts chose the scientific publications belonging to epidemiology, equity, vaccine, long-COVID and general COVID-19 issues. This gives us around 60 publications We used a web-based annotation tool[12] provided by deepset.ai to prepare this dataset in SQuAD 2.0 (Rajpurkar et al. 2016) format, which is a prototypical format for the question-answering task. We mark a snippet (a piece of text) from the full document as an answer (gold label) and formulate the corresponding scientific question. The gold label means a ground truth value (Hendrycks et al. 2018), which is an ideal predicted result based on humanly verifiable observation. After annotating, we get around 150 questions from these publications. We name this gold-standard dataset as **C**OVID-19 S**QuAD** (CQuAD) data, based on its data format on SQuAD. Besides our own created dataset, we also the following two benchmark datasets:

*B) COVID-QA (Möller et al. 2020):*

[12] https://annotate.deepset.ai/

COVID-QA[13] (**C**OVID-19 **Q**uestion **A**nswering) is a question-answering dataset consisting of 2,019 question/answer pairs annotated by volunteer biomedical experts on scientific articles related to COVID-19. This dataset is also made available in SQuAD 2.0 format that we need to use in a question-answering task of CO-SE architecture as well to evaluate the system performance.

*C) BioASQ (Tsatsaronis et al. 2015)*

BioASQ (**B**iomedical **S**emantic **I**ndexing and **Q**uestion **A**nswering) organizes challenges on biomedical question answering tasks. They recently release BioASQ Task Synergy for the COVID-19 question-answering task. The articles in the dataset are taken from CORD-19 and questions are provided by the experts as well as from the Trec-COVID-19 (Voorhees et al. 2020) challenge. The BioASQ task requires that the answers should be in one of the following formats: List; Yes/No; or Factoid. We chose to provide a list of at most top-k relevant articles/documents (the Retriever) as well as a list of at most top-k relevant text snippets as answers (the Reader) in these experiments (based on the research objectives). We use the latest train and test sets[14] for the Synergy task to train the reader component of CO-SE and to evaluate our approach respectively. This dataset is also available in SQuAD format, however, we perform additional processing to make it compatible with the SQuAD 2.0 format as required by our reader module.

[13] https://github.com/deepset-ai/COVID-QA
[14] http://participants-area.bioasq.org/Tasks/synergy/

## 4. Proposed COVID-19 Search Engine - CO-SE Architecture

We present the architecture of our proposed CO-SE framework in Figure 4 and explain its workflow below.

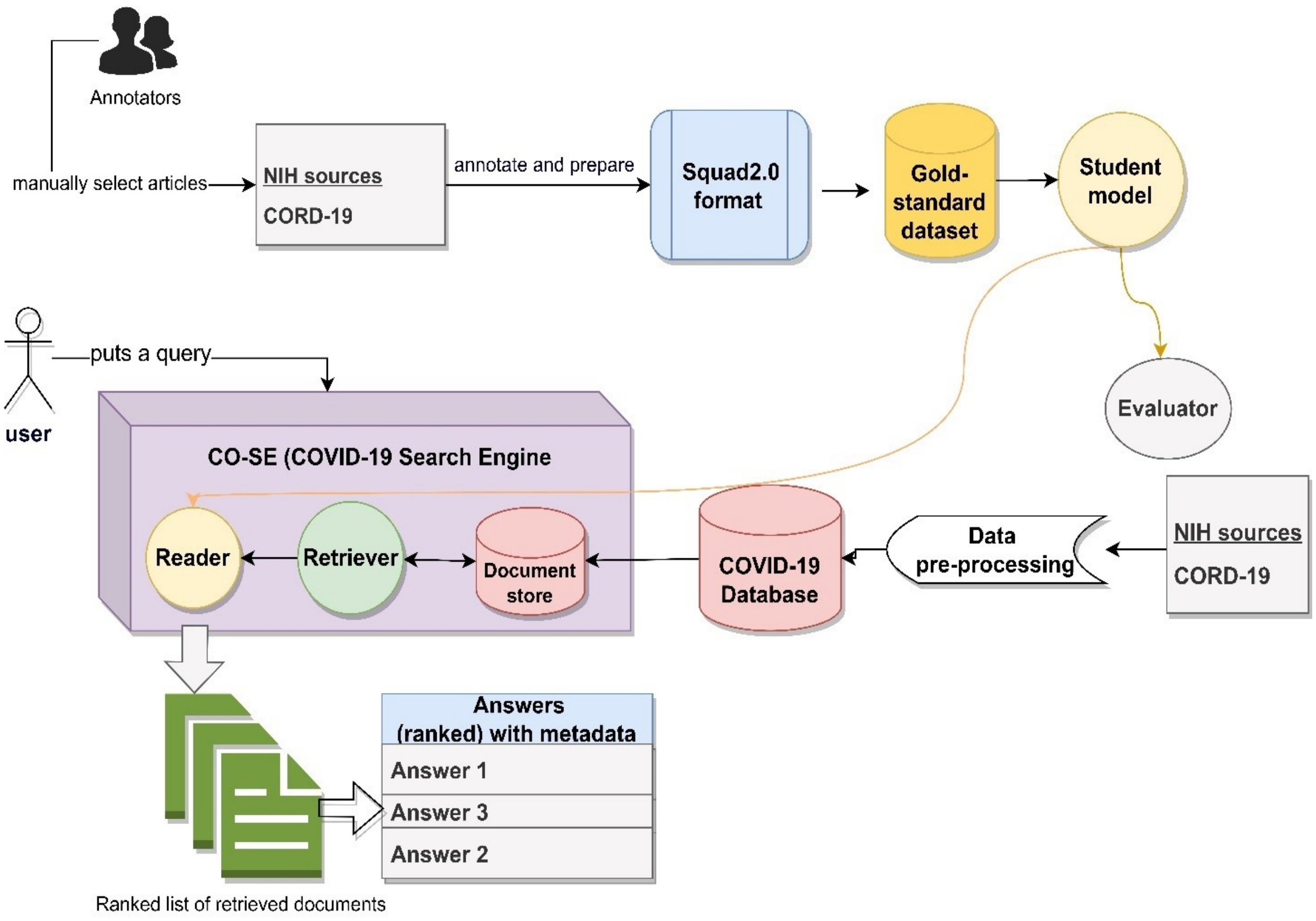

Figure 4: Overall architecture of CO-SE and its workflow

As shown in Figure 4 (right side), first, we get the publications data from the National Institute of Health (NIH)[15] source. We use the CORD-19 (Lu Wang et al. 2020) initiative to get the scientific publications related to COVID-19. Once we get the publications data, we prepare two datasets: (i) a COVID-19 dataset (shown as a pink cylinder in Figure 4); each scientific article from this dataset is stored in a database and sent to the CO-SE pipeline; and (ii) a gold-standard dataset (shown as a dark yellow cylinder in Figure 4); we use this dataset to evaluate our approach and to enhance the readability of the model. We also use the other two

[15] https://www.nih.gov/

benchmark datasets (COVID-QA and BioASQ) in this work. These datasets are discussed in Section 3.

### 4.1 Knowledge distillation from a teacher to a student model

Knowledge distillation is the process by which knowledge is transferred from a teacher (a larger) model to a student (a smaller mimic) model (Tang et al. 2019). We use the Bidirectional Encoder Representations from Transformers (BERT) (Devlin et al. 2018) as a ‘teacher’ model and the DistilBERT fine-tuned on our CQuAD dataset as the ‘student’ model.

*BERT as teacher model*: BERT is a Transformer-based architecture, which has demonstrated outstanding results in a wide range of NLP tasks (Rogers, Kovaleva, and Rumshisky 2020), including question answering (SQuAD v1.1 and 2.0), natural language inference, classification and related tasks. BERT is pre-trained on huge datasets (Wikipedia and Toronto Book Corpus) to achieve a large-scale language understanding.

BERT model can look at the words that come before and after a word to determine its full context, which is particularly useful for determining the intent behind a query. It employs the Transformer (Vaswani et al. 2017) model that has an attention mechanism to discover contextual relationships between words (or sub-words) in a text. A Transformer, in its simplest form, consists of two distinct mechanisms: an encoder that reads the text input and a decoder that generates a prediction for the task. Since BERT’s main objective is to generate a language model, only the encoder mechanism is required. We are interested in the BERT model to read the text from the retrieved documents by jointly conditioning on both left and right contexts.

Our focus, in this work, is on the SQuAD task of BERT (shown in Figure 5), which is related to the question-answering system. For the SQuAD task, BERT is fine-tuned on the SQuAD dataset. The original SQuAD dataset (Rajpurkar et al. 2016) is a crowd-sourced question-answering dataset with questions on Wikipedia articles and answers as text from the corresponding reading passage (contexts) or the questions are unanswerable. The SQuAD task requires a model to receive a query about a text sequence and mark the answer in the sequence. The BERT model for the SQuAD is then trained by learning two extra vectors that mark the beginning and end of the answer (as shown with red arrows in Figure 5).

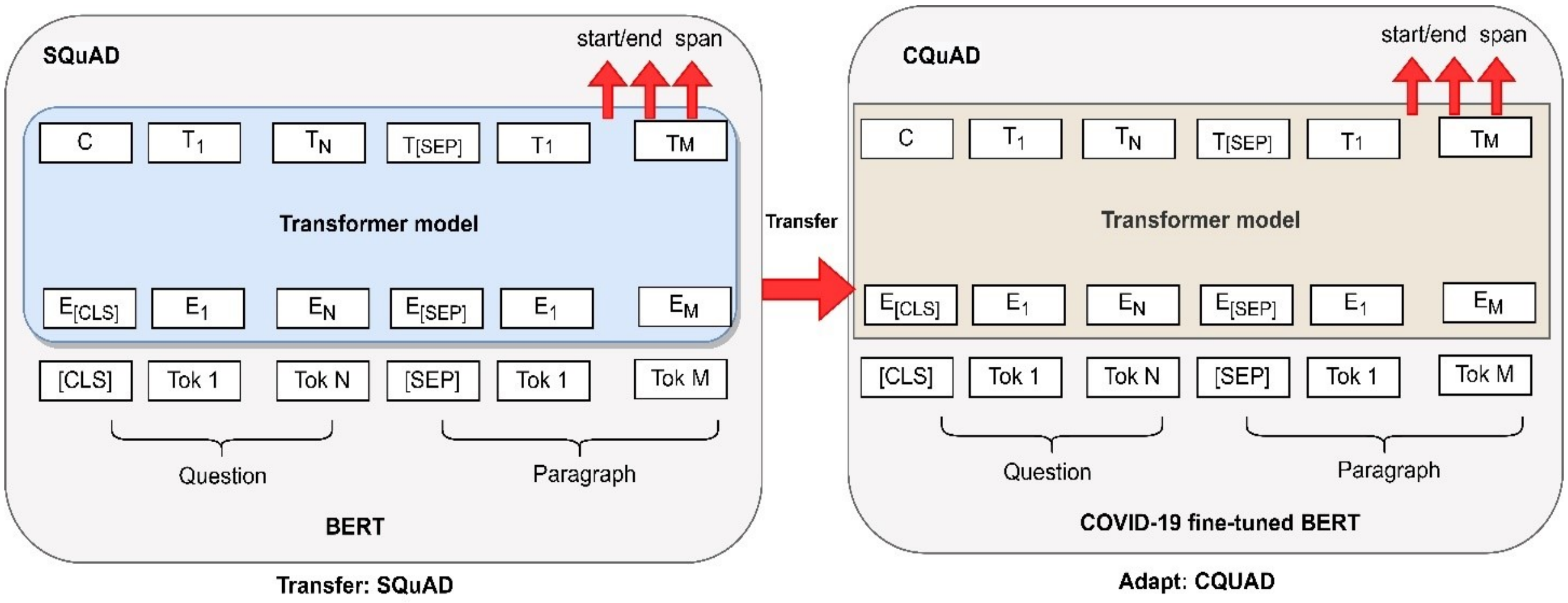

Figure 5: Adapting BERT for the SQuAD task for CQuAD data

*Distillation process*: The actual size of BERT is too large, which requires a lot of computational resources and makes it hard to scale well with the increasing dataset size (Morris 2020), so we utilize the distillation process of BERT. Distillation is a procedure for model compression, in which a smaller (student) model is trained to match a large pre-trained (teacher) model (Sanh et al. 2019). The first model to distil BERT is DistilBERT (Sanh et al. 2019), followed by TinyBERT (Jiao et al. 2020) and MobileBERT (Z. Sun et al. 2020). We are using the distillation process in this work for two reasons: it is a simple technique, produces good results, and allows using the BERT-based models with less computational resources.

As shown in Figure 6, the original BERT (teacher) model is fine-tuned on the SQuAD task. The BERT's encoder first gets the input embeddings of the SQuAD dataset and generates the output embeddings and provides a fine-tuned BERT (shown in a light blue block in Figure 6). In the next step, this fine-tuned BERT is then used to teach a student model. We use the distillation process of DistilBERT. In that, we take the distilbert-base-uncased with 6 layers, 768 dimensions and 12 heads, totalizing 66M parameters to create a student model (shown in a light-yellow block in Figure 6). We use the DistilBERT code from here[16] and fine-tune it for our CQuAD dataset (shown as a dark yellow cylinder in Figure 6) to produce our fine-tuned student model (shown in a light purplish block in Figure 6) with the distillation loss function.

[16] https://github.com/huggingface/transformers/tree/main/examples/research_projects/distillation

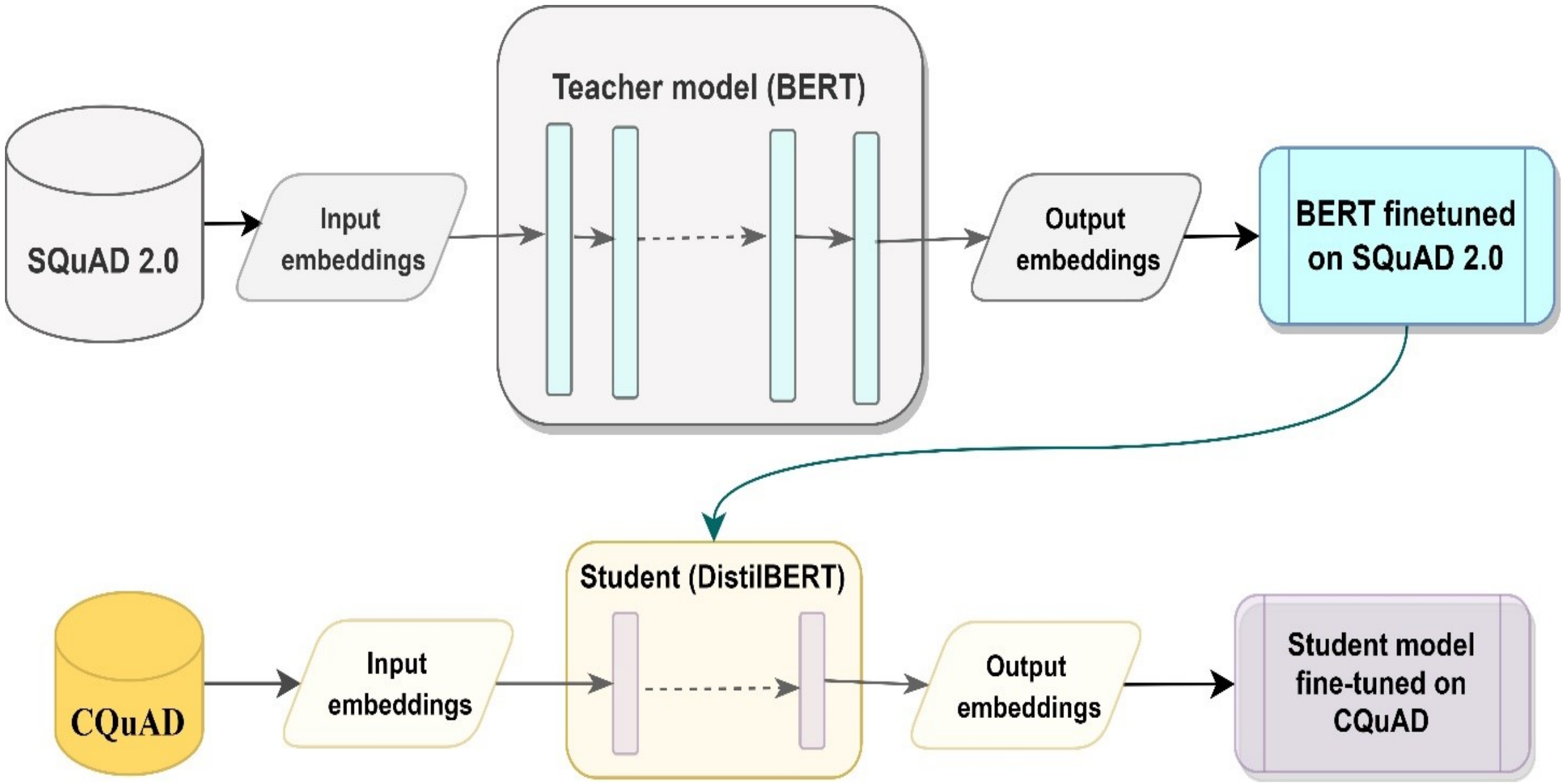

Figure 6: Distillation process of BERT (teacher) to our student model

We have released the model weights of our student model here[17], which can be used to build a search engine or question-answering system related to COVID-19 by using the following line of code.

```
git lfs install
git clone https://huggingface.co/shaina/covid_qa_distillBert
```

Our intuition behind this whole fine-tuning process is that it is computationally less expensive than pre-training the model from the scratch (Jiao et al. 2020). Also, by fine-tuning it on our gold-standard dataset, we modify only the outer layer of the main model (DistilBERT) to recognize whether a question is answerable or not (SQuAD) task.

### 4.2. CO-SE Pipeline

The main part of this proposed architecture is the CO-SE pipeline, which consists of a retriever component, a document store, and a reader component. Since this pipeline is a part of the overall CO-SE architecture (Figure 4), so we call it a CO-SE pipeline. We propose the design of our CO-SE pipeline in Figure 7 and explain its work below.

[17] https://huggingface.co/shaina/covid_qa_distillBert

As shown in Figure 7, a user gives a query related to COVID-19 that goes to the CO-SE pipeline. Inside the CO-SE pipeline, the query goes to the retriever module. The retriever searches for the relevant documents related to a search query from a document store that stores COVID-19 articles (each article here is a document). The retrieved set of documents from the retriever then goes to the reader, which finds the answers from the retrieved documents. The output from the CO-SE is a ranked list of search results (as documents) presented to the user. Each result consists of an article title, an answer, a context (surrounding paragraph around the answer) and other metadata (DOI of article, authors, journal details) information. next, we explain each component of the CO-SE pipeline below:

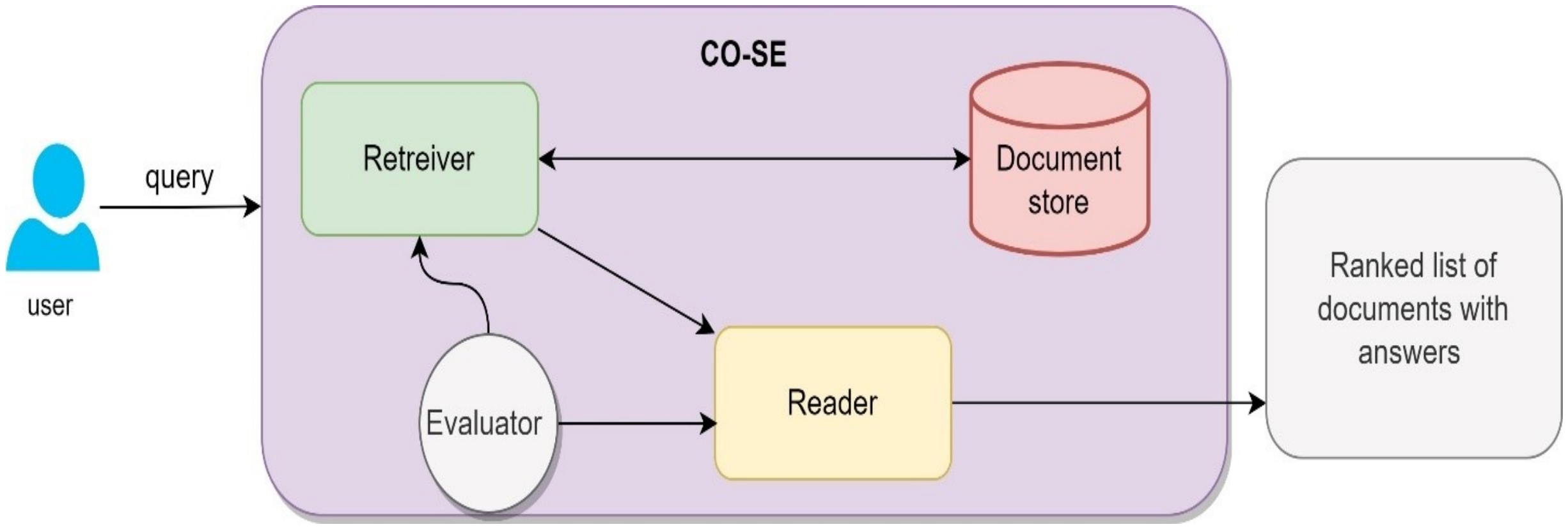

**Figure 7**: CO-SE pipeline

*Retriever*: The retriever is a component within the CO-SE pipeline that retrieves a set of documents from the document store that it determines to be most relevant to a query. It acts as a filter that finds the best candidate documents by calculating the similarity between the question and the documents. In this work, we use the Term Frequency-Inverse Document Frequency (TF-IDF) model (Aggarwal 2015) in the retriever component.

The TF-IDF calculates the inverse proportion of a word in a single document to the inverse proportion of that word across the entire corpus of documents. The basic idea behind using the TF-IDF algorithm within the retriever component is to convert the calculation of a given query similarity into the calculation of angles between documents. The calculated scores are then used to retrieve documents most relevant to the given query. The retrieved documents are also displayed in relevance order (the highest score corresponds to the most relevant document and so on).

*Document store*: The document store is like a database that takes the scientific articles from our COVID-19 database and index each document. We use Apache Tika[18] for parsing the full texts and to extract metadata (title, authors, DOI, publication date) from the articles. We also clean the text and split long articles into multiple smaller units as part of the preprocessing. The converted data then goes into the document store. We refer to an individual piece of article stored in the document store as a document. We store the information – text and metadata – corresponding to each document as data frames (i.e., a columnar field for each piece of information -title, text, etc.,).

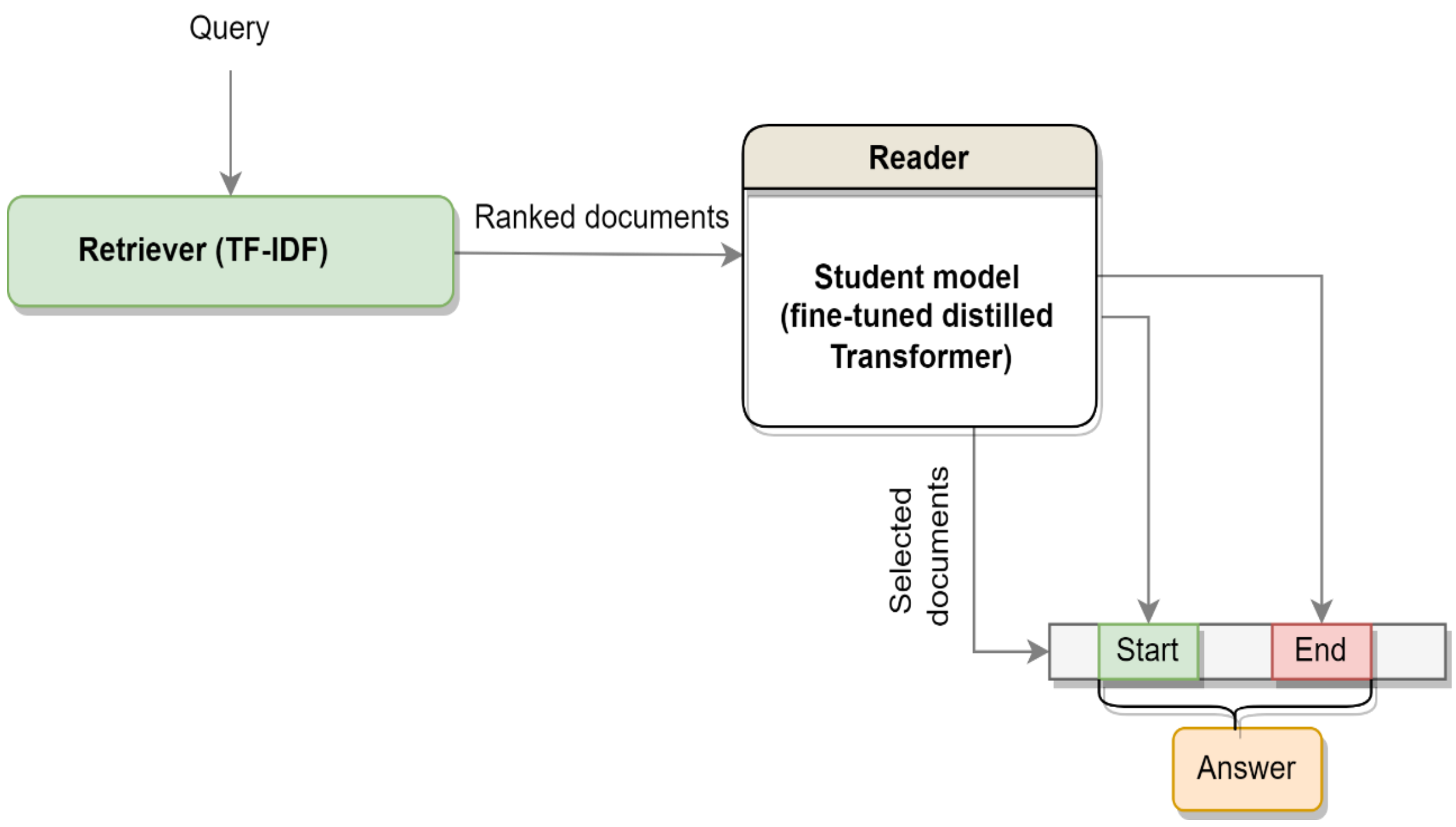

Figure 8: Reading comprehension of reader component

*Reader*: The reader provides a ranked list of answers based on the query being asked. The input to the reader is a set of documents that are ranked by the retriever and the output is a list of answers for each query. The reader provides the answer and a context from the paragraphs within the documents in response to each query. In this work, we use our student model

[18] https://tika.apache.org/

(shown in Figures 5 and 6) inside the reader component to transfer knowledge from the powerful neural networks into our specialized task.

The pre-trained language models (Bert, RoBERTa and others) have already been tested to be quite effective at question-answering tasks (Pearce et al. 2021), which motivates us to use one such model in our reader component. We use our student model (Transformer model) to take advantage of deep neural networks' ability to read through texts in detail to find an answer. Our goal is to determine whether an answer to a question exists in a given context, thereby encouraging the development of reading comprehension models with a better understanding of language. Because the CO-SE document store is too large and will continue to grow with upcoming COVID-19 publications, so we prioritize speed and GPU memory over higher accuracy in this study, therefore, we use the distillation process of a larger pre-trained model (shown in Figure 6). We use three datasets to train our reader module, which is: (i) CQuAD; (ii) COVID-QA; and (iii) BioASQ.

Our reader model differs from other Transformer-based question-answering systems (many of which can also be found on Huggingface.co) in that other model require explicit contexts (paragraphs) to provide an answer. In contrast, our reader component does need any explicit contexts. The reader in CO-SE gets its contexts from the retrieved documents provided by the retriever in a pipeline, as shown in Figure 8. The reader then returns the answer(s) with a supporting context. It also provides the start and end position of text from the retrieved documents as the context, along with metadata information.

*Answers:* The output of the CO-SE pipeline is a ranked list of documents based on the query asked by a user. Each document is accompanied by an answer, a context or a paragraph from which the answer is extracted. The reader also shows the model's confidence in the accuracy of the extracted answers.

*Training and evaluation*: Both the retriever and reader are chained together in the CO-SE pipeline. This pipeline is represented as a directed acyclic graph of component nodes. It enables custom query flows, merges candidate documents for a reader from the retriever, and performs re-ranking of candidate documents. The purpose of CO-SE architecture is to automate different steps of the model life cycle, starting with data ingestion, data pre-processing, and model training to provide answers. The CO-SE pipeline (Figure 7) gets its input from our COVID-19 dataset, on which the pipeline is then trained.

This CO-SE pipeline includes the raw data input, features, outputs, machine learning models (retriever and readers), model parameters, and prediction outputs (answers). We also have two evaluation components: one inside this pipeline (shown in a grey circle in Figure 4) to evaluate the performance of individual components (retriever and reader), and the other outside the CO-SE pipeline to evaluate the performance of the student model (shown in the grey circle in Figure 7).

### 4.3. Toy example of the query with answer and metadata

In this section, we demonstrate the working of CO-SE with a toy example.

As a first example, shown in Figure 9, we enter a query '*What are symptoms of COVID-19*?". We specify top@ 10 for the retriever to retrieve the documents. For brevity reasons, we specify top@ 2 for the reader to return us the top-2 ranked documents with answers. In response to CO-SE, we get the answers: "*tearing of the eyes, sore throat, cough, and runny nose*" with 96.78% model's confidence, and "*wheezing episodes, asthma exacerbations, acute obstructive bronchitis, bronchiolitis, croup, and pneumonia*" with 88.5% score. Our CO-SE pipeline also shows the metadata and confidence of the model for each query.

| index | answer | type | score | context | meta | offsets in document | offsets in context |
|---|---|---|---|---|---|---|---|
| 0 | tearing of the eyes, sore throat, cough, and runny nose | extractive | 0.967710733 | om residents with recently installed urea-formaldehyde foam insulation who complained of formaldehyde odor, irritation, or increased pre-existing illness patterns. In 40/55 homes investigated, the most common symptoms were tearing of the eyes, sore throat, cough, and runny nose. Air samples were collected in 22 homes. In 14 homes where formaldehyde was detected, levels ranged from 0.01 to 0.78 ppm.In New Hampshire, | {'name': 'Report of the Federal Panel on Formaldehyde*'} | {'start': 97127, 'end': 97182} | {'start': 223, 'end': 278} |
| 1 | wheezing episodes, asthma exacerbations, acute obstructive bronchitis, bronchiolitis, croup, and pneumonia | extractive | 0.885707617 | Koch are not applicable. 10 Numerous retrospective and prospective studies suggested an association between HBoV and acute respiratory tract infections in children. Clinical fi ndings consisted of wheezing episodes, asthma exacerbations, acute obstructive bronchitis, bronchiolitis, croup, and pneumonia. [11] [12] [13] [14] [15] Infants below the age of two years and patients with structural pulmonary diseases or | {'name': 'Frequency and clinical relevance of human bocavirus infection in acute exacerbations of chronic obstructive pulmonary disease'} | {'start': 1874, 'end': 1980} | {'start': 197, 'end': 303} |

**Figure 9**: Metadata with each result.

The answers returned by CO-SE are extractive. Extractive question-answering is the process of looking through a large collection of documents to extract a concise snippet to answer a question (P. Lewis et al. 2019). As our reader is based on the SQuAD task, so the reader also returns the offset (start and end position of words) of the answer in the whole document as well as in the context. It also returns the metadata (title in this example) and the model's confidence score on the answer.

## 5. Experimental Setup

In this section, we discuss our experimental setup.

### 5.1. Settings and Hyperparameters

For training, we used an Nvidia Tesla P100 GPU with 16 GB RAM and 2TB disk storage. We use TensorFlow as our deep learning framework and Python as our programming language. We test various values for the hyperparameters and are reporting the optimal values below:

- The total batch size for training is set to 16. We use a smaller batch size to fit the training data in the given memory.
- The maximum query length is set to be 100 tokens (words).
- The maximum answer length is set to be 250.
- The maximum sequence length, which is the size of the input document sequence, is set as 512.
- The pre-trained word embeddings dimension that is used is 768.
- The Adaptive Moment (Adam) estimation (Kingma and Ba 2015) optimizer is used with a learning rate of 1e-8. In addition, the $L_2$-regularization and dropout methods are included in the training process to avoid the problem of over-fitting.
- The $L_2$-regularization is set to be 1e-5 and the dropout ratio is set to be 0.75.

For all questions and answers, if the sentence length exceeds or falls below the required length, we pad or truncate it. We train our models in mini-batches and use the exponential decay method to vary the learning rate in each epoch, with a decay rate of 0.9. Each experiment is repeated at least 10 times. All the baseline models are also optimized to their optimal settings, and we also report the best result for each baseline model.

## 5.2. Baseline Methods for Comparison

We use the following baseline methods to compare against our model.

*BERT* (Devlin et al. 2018): Bidirectional Encoder Representations from Transformers (BERT) is a Transformer-based model pre-trained using masked language modelling objective and next sentence prediction on a large Wikipedia corpus. In this work, we use the BERT-Base, Uncased, which has 12 layers (transformer blocks), 12 attention heads, and 110 million parameters.

*BART* (M. Lewis et al. 2019): Bidirectional and Auto-Regressive Transformer (BART) is a Transformer-based model that uses a standard sequence-to-sequence architecture. We use the BART-Base model with 12 layers (6 encoder and decoder layers) and 217 million parameters in this work.

*XLNET* (Z. Yang et al. 2019): XLNet is a Transformer-based model pre-trained on generalized permutation language modelling objective. We use XLNet-Base, Cased with 12-layers, 12-heads and 110 parameters as a baseline model.

*LongFormer* (Beltagy, Peters, and Cohan 2020): Longformer is a modified Transformer that can process long sequences and scales quadratically with sequence length. We use Longformer-base having 12-layer, 12-heads, and around 149M parameters.

*Funnel* (Dai et al. 2020) A Funnel Transformer is a type of Transformer that gradually compresses the sequence of hidden states to make it shorter, lowering the computation cost. In this work, we use the Funnel-transformer-small version having 14 layers, 12-heads and 130 million parameters.

*COVID-QA system* (Möller et al. 2020): COVID-QA is a question answering system based on the Robustly optimized BERT approach (RoBERTa) model (Liu et al. 2019). RoBERTa is the retraining of BERT with improved training methodology, more data and better computations.

We report the results for each baseline according to its optimal hyperparameter setting and report the best results for each baseline. We have divided the COVID-19 dataset into training, validation, and test sets, with a 75:15:15 ratio for all experiments. We evaluate the results using CQuAD, COVID-QA and BioASQ evaluation datasets.

### 5.3. Evaluation Metrics

In this work, we make use of the following evaluation metrics:

- Precision (prec), Recall and Mean Reciprocal Rank (MRR) to evaluate retriever.
- Accuracy (acc), Exact Match (EM) and Semantic Answer Similarity (SAS) to evaluate reader.

*Precision* is the fraction of retrieved documents that are relevant (Teufel 2007).

*Recall* is the fraction of relevant documents that are retrieved (Teufel 2007).

*Mean Reciprocal Rank (MRR)* is a relative score that calculates the average of the inverse of the ranks at which the first relevant document is retrieved for a set of queries (Teufel 2007).

*Exact match (EM)* measures the proportion of documents where the predicted answer is identical to the correct answer (Rajpurkar et al. 2016).

*Accuracy* is defined as the proportion of correctly classified items, either as relevant or as irrelevant (Teufel 2007).

*Semantic Answer Similarity (SAS)* (Risch et al. 2021) metric takes into account whether the meaning of a predicted answer is similar to the annotated answer, rather than just the exact words comparison. We employ a Transformer-based "cross-encoder/stsb-RoBERTa-large"[19] pre-trained model, to determine the semantic similarity of two answers.

We demonstrate the results of all models (our CO-SE and all baseline models) for various top@ k values. The top@ k refers to the number of relevant documents returned from the top-k retrieved documents. In this study, we use k values of 5, 10, and 20 for top@ k based on the established heuristics in IR evaluation (Schütze, Manning, and Raghavan 2008). Generally, a higher score on any of the above-mentioned metrics indicates a higher value.

## 6. Results and Analysis

In this section, we evaluate the results of our CO-SE pipeline. The goal of this evaluation is to see how well our model works in each setting and which module of the pipeline needs to be improved. We evaluate both the retriever and reader modules within the pipeline with different top @k values on our evaluation datasets (evaluation dataset details are in Section

[19] https://huggingface.co/cross-encoder/stsb-RoBERTa-large

3). Each module is evaluated based on its evaluation metrics. The results are shown in Table 2 and discussed in the following sections.

**Table 2**: Evaluation of CO-SE (bold means highest score) on all evaluation datasets

| **Datasets** | **Retriever** | | | **Reader** | | |
|---|---|---|---|---|---|---|
| | **Recall@5** | **Recall@10** | **Recall@20** | **EM@5** | **EM@10** | **EM@20** |
| **CQuAD** | **0.639** | 0.621 | 0.736 | **0.549** | **0.594** | **0.714** |
| **COVID-QA** | 0.620 | **0.724** | **0.824** | 0.544 | 0.536 | 0.700 |
| **BioASQ** | 0.626 | 0.624 | 0.724 | 0.519 | 0.519 | 0.699 |
| | **MRR@5** | **MRR@10** | **MRR@20** | **SAS@5** | **SAS@10** | **SAS@20** |
| **CQuAD** | 0.532 | 0.567 | 0.614 | **0.623** | **0.687** | **0.785** |
| **COVID-QA** | **0.640** | **0.713** | **0.750** | 0.620 | 0.662 | 0.769 |
| **BioASQ** | 0.512 | 0.622 | 0.692 | 0.438 | 0.513 | 0.595 |
| | **prec@5** | **prec@10** | **prec@20** | **acc@5** | **acc@10** | **acc@20** |
| **CQuAD** | 0.286 | 0.234 | 0.244 | 0.812 | **0.864** | **0.853** |
| **COVID-QA** | **0.321** | **0.281** | **0.260** | **0.832** | 0.840 | 0.870 |
| **BioASQ** | 0.318 | 0.257 | 0.224 | 0.802 | 0.823 | 0.824 |

Next, we discuss the performance of the retriever and reader.

### 6.1. Performance of Retriever

Table 2. displays the performance of the CO-SE retriever on different evaluation sets.

Overall, the results show that the retriever performs best on COVID-QA, then CQuAD, and finally BioASQ. This is probably because COVID-QA provides a complete set of question-answering pairs with full detail paragraphs (contexts) of 2,019 articles, so it is easy for the retriever to find many matching documents based on a given query. Our gold-standard CQuAD is smaller in size than COVID-QA, so it may have missed good hits during the retrieval process in the test phase. However, the difference in retriever's evaluation scores between the two datasets is negligible. We also notice that the CO-SE retriever performs slightly less when tested on BioASQ when compared to other datasets. This is most likely because BioASQ has a limited number of question-answering pairs and contexts are also not as extensive as in other datasets. As a result, the retriever may not be able to pick many relevant documents.

The results in Table 2 also show that as the value of top@ k increases, the recall and MRR scores improve (get higher close to 1). This is evidenced by the relatively high recall and MRR scores for the relevant documents during top @ 5, 10, and 20. A higher recall value indicates that our system can retrieve many truly relevant documents, in response to each query.

The precision score shows the number of relevant items that are returned. As shown in these results (Table 2), the retriever's precision decreases as top @k increases, and the overall precision score is lower than the recall score. Typically, as recall increases, precision decreases and vice versa (Schütze, Manning, and Raghavan 2008). In this work, we are more interested in determining the total number of relevant documents retrieved, and thus recall is a higher priority for our system.

We also see that the retriever's performance for MRR increases with increasing top@ k. This indicates that our method is quite accurate at locating the list's first relevant element. For example, if a search for a specific question "What is COVID-19?", returns a relevant document at the 1st position, its relative rank is 1, if the relevant document is at position 2, then the relevant rank is 0.5 and so on, and if there are no relevant documents then the score is 0. When the relevant ranks are averaged across the set of queries, this is the MRR. This measure is usually more appropriate for targeted searches, such as those in which users inquire about the first best item (Teufel 2007).

### 6.2. Performance of Reader

We evaluate the reader's performance based on how well it extracts the best answers from the documents retrieved by the retriever. We find (in Table 2) that the reader performs the best (overall) on our CQuAD dataset, followed by COVID-QA and then BioASQ. The improved performance of the reader on CQuAD is most likely due to the inclusion of all text and metadata information, which naturally assists the reader to query the data easier and recognize key information in response to each query. The reader's performance on COVID-QA is also very close to CQuAD; COVID-QA also includes the full text of the documents in the dataset, which helps to improve the readability of the model. However, because CQuAD is more versatile in terms of topics such as COVID-19 effects, equity, and long-COVID, despite its small size, it provides more readability than other datasets, as demonstrated in these experiments. In addition, CQuAD also includes full texts and metadata that enhance the reading performance of the reader. The lower performance of CO-SE on BioASQ is most likely due to the limited information compared to other datasets, which affects the reader's reading comprehension in retrieving semantically accurate information.

The results in Table 2 also show that our reader is quite good at returning correct answers. This is demonstrated by the reader's accuracy score of more than 81% on all datasets. The EM ratio of the reader also increases with top@ k (it is about 70% during top@ 20 on all datasets) showing the ability of this component to give a precise answer.

Normally, the SAS is a critical metric for reading comprehension task (Rajpurkar et al. 2016). The SAS score of our Reader is also above 76% during top@ 20 for CQuAD and COVID-QA datasets. This result shows the high semantic textual similarity between the predicted and the ground truth answer.

*The general takeaway from the results:* In our retriever' and reader's results, we see that we get better performance for most metrics with increasing top@ k. Though there is no rule of thumb that increasing top@ k improves accuracy, mostly the experimental results in IR works (Bai et al. 2005; Schütze, Manning, and Raghavan 2008; Dacrema, Cremonesi, and Jannach 2019) show that increasing top@ k generally improves the model performance (in terms of accuracy). This is self-evident because a system that retrieves the relevant documents at higher ranks and returns a greater number of relevant docs would score higher than a system that fails to satisfy either or both cases.

In this work, we use the value of top@ k till 20 based on general heuristics in the evaluation of such systems (question-answering, recommender systems) (F. Sun et al. 2019; Vargas and Castells 2011; Bai et al. 2005). While it is true that commercial search engines such as Google or Bing return many results in response to a single query, however, we reason that most users do not have the patience to go beyond 10 or 20 retrieved items (waiting and scrolling the results). So, to evaluate the results, we only include till top@ 20. However, we can easily increase the number of retrieved results to 100, 1000, or more if it is the goal.

### 6.3. Performance comparison with baselines

We also compare the performance of state-of-the-art systems to our CO-SE system on three datasets (CQuAD, COVID-QA and BioASQ). The primary goal of this set of experiments is to see how well different models (including our and other baseline models) perform. Since these baselines do not necessarily have the same architecture as ours, so we evaluate the ability of each system based on its reading comprehension, i.e., how well it performs while providing answers in response to a query (reading is a common phase in all these baselines and our method). We use the SQuAD evaluation standard (Rajpurkar et al. 2016), i.e., EM and SAS metrics, to assess each system's ability in providing answers to a query from the retrieved documents. We report the results of all models during top@ 20, based on the best results in earlier experiments (Table 2). The results are shown in Figures 10 and 11 and discussed next.

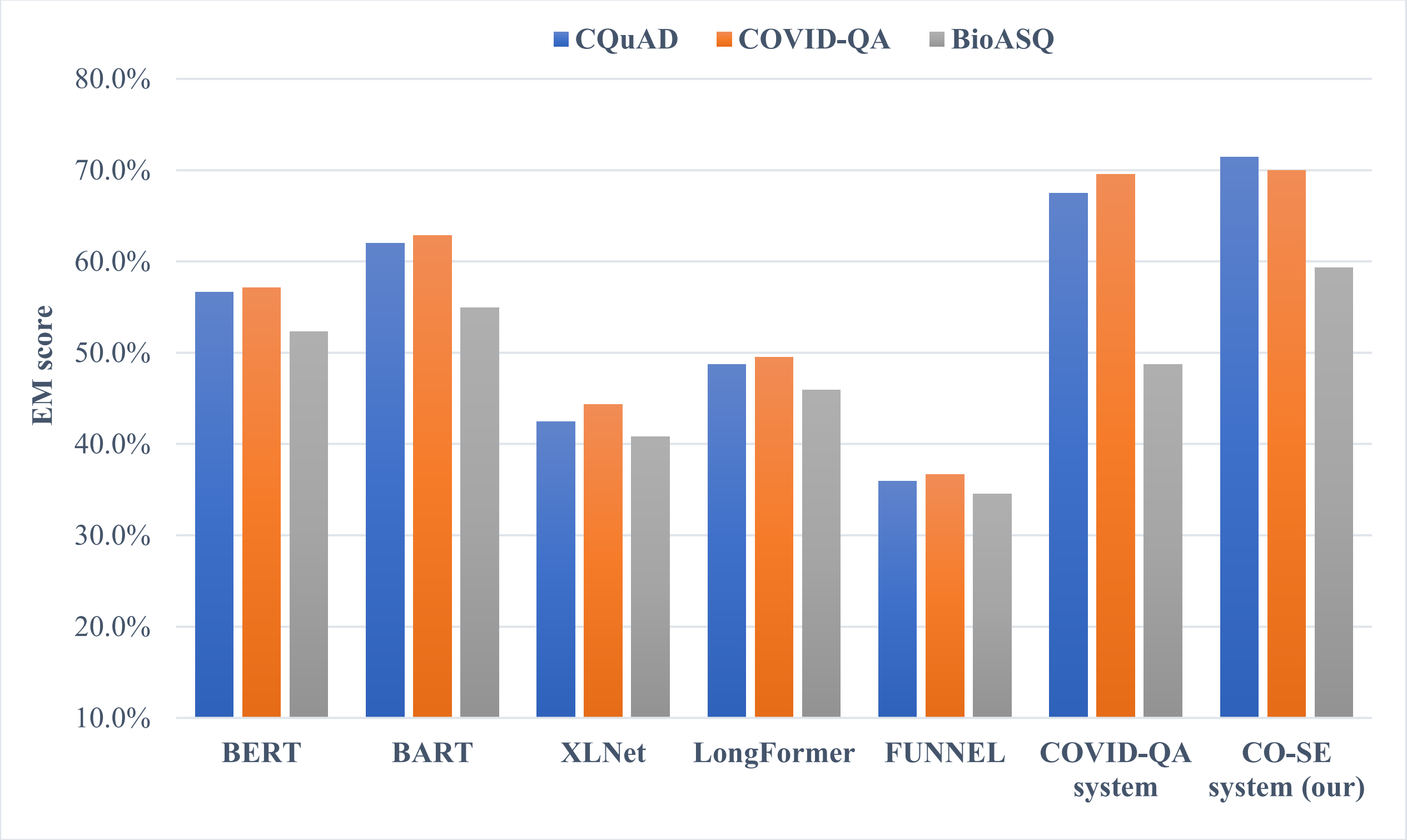

Figure 10: Comparing CO-SE and baselines for EM scores on reading comprehension using all datasets

As shown in Figures 10 and 11, our CO-SE model outperforms all baseline models for EM and SAS scores. This is demonstrated by our model's highest EM score of 71.45 % on CQuAD, 70% on COVID-QA and about 60% on BioAS; and highest SAS score of 78.5% on CQuAD, 76.90% on COVID-QA, and about 60% on BioASQ. This demonstrates that our model outperforms other models in all dataset settings.

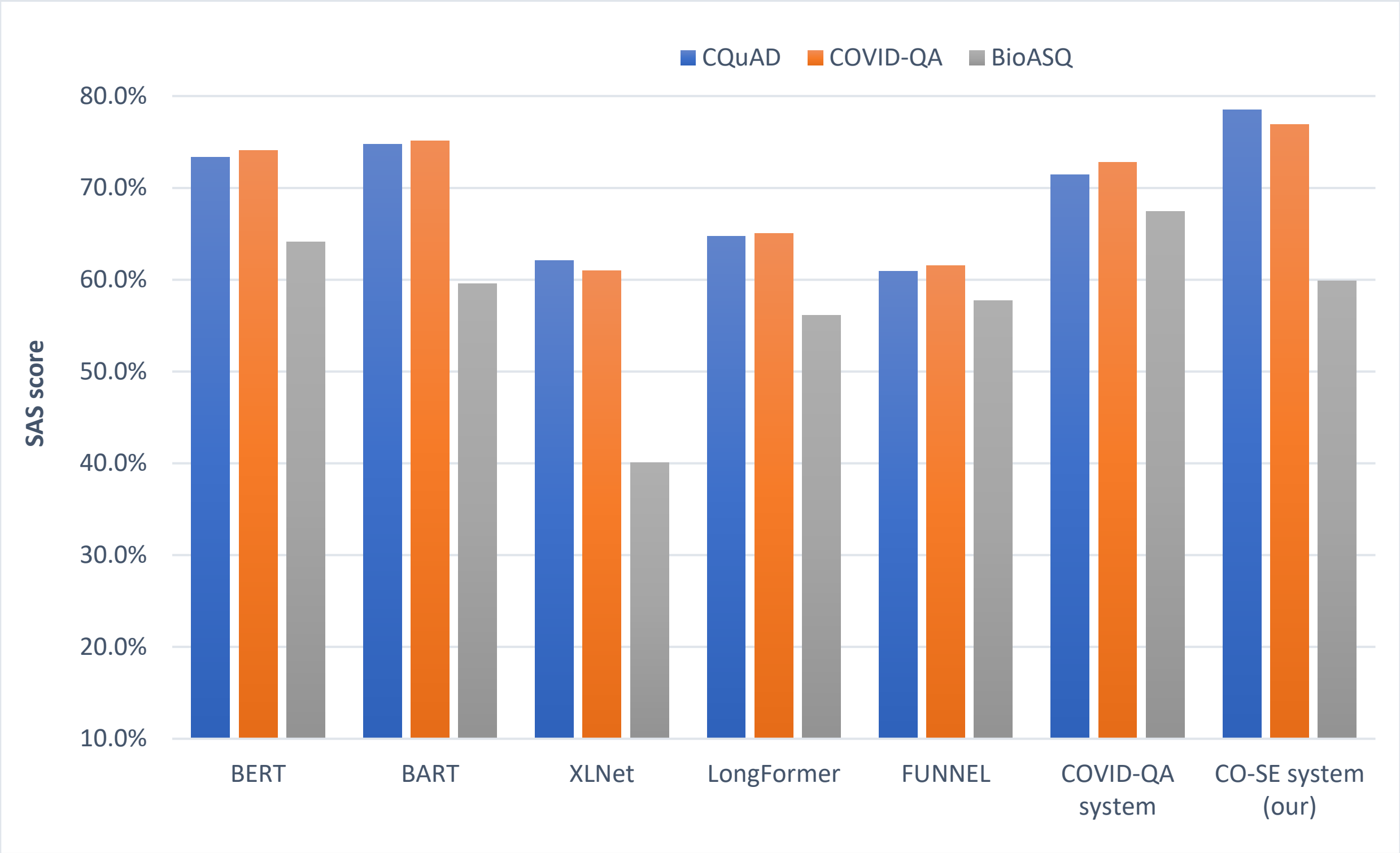

Figure 11: Comparing CO-SE and baselines for SAS scores on reading comprehension using all datasets

We observe in Figures 10 and 11 that the CO-SE reader performs better compared to all baseline methods. Among the three datasets, CO-SE performs better when we test it on CQuAD (our dataset), followed by COVID-QA and BioASQ (benchmarks). This also shows the generalizability of our approach across different types of COVID-19 topics, this is probably because compared to other datasets, CQuAD covers a wide range of topics.

The performance of the COVID-QA system (Möller et al. 2020), which is based on RoBERTa, is next. We can see that the COVID-QA system performs quite well on the COVID-QA dataset (on which it was originally trained), followed by CQuAD and then BioASQ. The lower performance of the COVID-QA system on CQuAD is most likely because CQuAD contains some question-answering pairs (e.g., vaccine, long-COVID, equity) that are not present in the COVID-QA dataset, so a drop in performance is to be expected. However, the performance difference between the COVID-QA system on both the datasets (COVID-QA and CQuAD) is not quite significant. This could imply that RoBERTa is a good candidate model for our reader component. The RoBERTa model strengthens the COVID-QA model to retrieve the documents and read the answers accurately and efficiently.

Then, in the same order, BART and BERT perform. BART also performs well in general SQuAD tasks and can handle sequences of up to 1024 tokens (Liu et al. 2019). BERT's SAS metric performance is quite good (around 74% on CQuAD and COVID-QA). This demonstrates the BERT's ability to provide a semantically correct answer. In terms of model performance on datasets, we see that these models perform nearly equally on CQuAD and COVID-QA (except for some places, where they perform better than each other marginally). The BioASQ dataset has a low impact on model performance, which we believe is because it lacks the detailed contexts, paragraphs, and metadata that the other two datasets do.

Next comes the performance of Longformer, XLNet and Funnel in the same order. The advantage of Longformer compared to BERT is that it can handle trained sequences of text. For example, Longformer can handle a text of 5000 words or up, which is normally a length of a publication or scientific article. However, in our experiments, BERT has shown better performance than Longformer, since we are able to fit all the data into the memory by using the proper batch sizes. We also observe that the masked language modelling and next sentence prediction task of BERT performs better than XLNet with the permutation language modelling in the reading task. We also observe that the Funnel performs an average for reading comprehension in our experiments. One may benefit from this model if there are limited resources (memory, disk, CPU cycles) and the goal is to perform a natural language understanding task.

### 6.4 Effectiveness of student model

In this experiment, we evaluate the performance of our student model, which is a Transformer based model and a fine-tuned version of DistilBERT on CQuAD. We show the training and validation loss where the validation set is 15% of the training set. We set the optimizer to Adam trained for 50 epochs (after so many epochs, there are normally not any improvements that we also observe, so we report the results with 50 epochs). We use the loss function as defined in the SQuAD paper (Rajpurkar et al. 2016), which is the sum of cross-entropy loss for starting word and ending word positions. The purpose of this experiment is to check the effectiveness of our student model for the best fit of line.

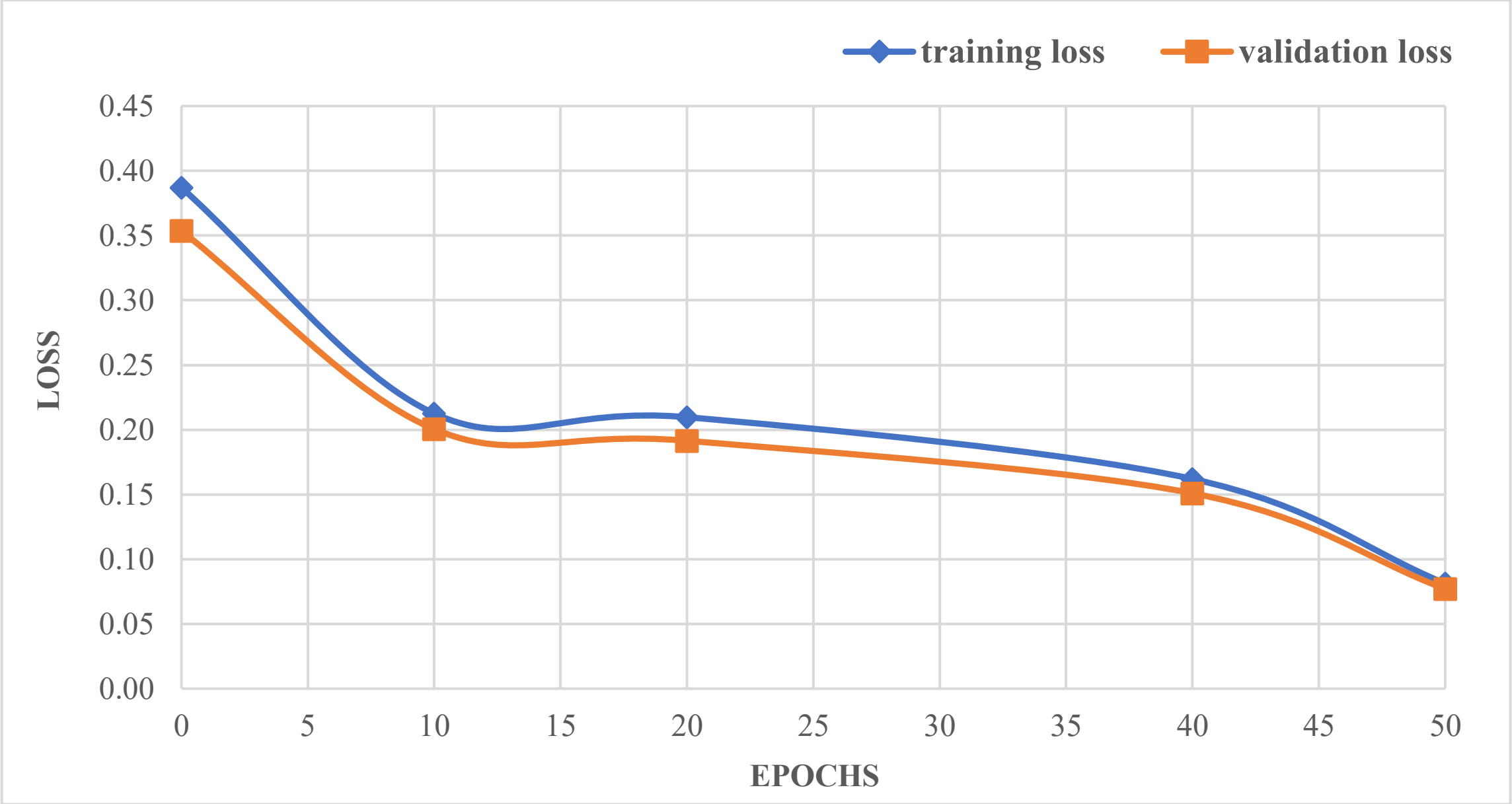

Figure 12: Effectiveness of student model

As shown in Figure 12, both the training loss and the valid loss are decreasing with every epoch. This shows that we are achieving a good fit of our model on the training data, and it also generalizes well on new, unseen test data. This is consistent with some of the findings in transfer learning literature (Rajpurkar et al. 2016; Weiss, Khoshgoftaar, and Wang 2016; Devlin et al. 2018) which has shown that a carefully fine-tuned model with rich linguistic information can easily adapt to new domains (which is COVID-19 in this work).

## 7. Discussion: limitations and future perspectives

In this section, we discuss the practical impact, limitations of the data and methods.

### 7.1. Practical Impact

Thousands of documents are being produced every day during this period of crisis during COVID-19, but only a small percentage of them are scientific, rigorous, and peer-reviewed. This may result in the inclusion of incorrect material as well as the possible quick dissemination of research and data that is scientifically disprovable or otherwise wrong (Raza and Ding 2022). In this time-sensitive environment, those on the front lines - such as medical practitioners, policymakers, and other decision-makers - may be unable to process the torrents of scholarly literature on COVID-19. The ability to search for COVID-19 specific information is critical to making this plethora of information both meaningful and actionable. Curation of scientific publications can also be used to offer the most frequently asked queries by the medical research community at a given time. We develop CO-SE with this goal in mind to help researchers find useful and credible scientific information on time.

CO-SE is an important step toward assisting medical researchers in quickly and meaningfully locating relevant content. We have offered the design of an architecture for researchers in the field of AI to construct such a system and assist the health science community in fighting against the pandemic by getting the most up-to-date research and findings in less time. We also released the weights of our fine-tuned student model on the CQuAD data that may be utilized for a variety of tasks such as COVID-19 question answering, COVID-19 article summarization, and translation. Such a system can be used in the healthcare setting, where doctors, nurses and scientists can use it to get up-to-date scientific information on time and can help mitigate the pandemic effects or to address other disease issues as well, on time.

### 7.2. Reusability of pipeline and generalizing to other domains

Our CO-SE architecture is flexible and adaptable, the pipeline allows reusability of models by allowing for rapid and flexible extensions to different intervention scenarios. This architecture is also updatable with the evolving literature on the disease (e.g., COVID-19 or its variants). Besides the COVID-19 scenarios, this architecture can be adapted to other health science fields, such as searching for the relevant literature related to epidemiology, learning health systems, and different biomedical or pure medicine use cases. The only pre-requisite is to update the data source, and parameters and to perform additional fine-tuning. For example, our student model is adapted from the BERT model, if the data source is related to studying diseases from medical documents, or studying clinical case reports, then the student model should be fine-tuned on the relevant data. Additionally, we design this architecture to be tailored to other domains (e.g., journalism, student learning systems, and so on) that are not limited to health sciences or COVID-19. This architecture can be a useful tool for public health decision-making when used correctly as part of an iterative decision-making process.

### 7.3. A design approach

CO-SE is a design strategy that has been proposed at the academic level. The goal of this research is to show how we can create a search engine that can be used for mining biomedical data. It is suggested that this approach be used with an interface that allows for real-time searching, which is currently a limitation of the system. It is, therefore, recommended that when using an interface, searchers begin with one of those databases (such as CORD-19 or LitCOVID) to develop and fine-tune search strategies.

Our goal, in this research, is to propose the design of a search engine that addresses the shortcomings of current biomedical portals, such as the fact that proximity searches on the PubMed interface must be conducted using phrases or Boolean “AND” conjunctions. Using Boolean “AND” combinations and phrase searching complicates the process and increases the likelihood of missing relevant articles. However, when using our approach, users only need to enter a query, and the proposed design handles the intermediate operations. The methods, described in this work, can be used to create complex and comprehensive search strategies for various other databases, such as those required when searching for relevant references for systematic reviews. Such a system, if implemented, can help both information specialists and practitioners when searching the biomedical literature.

### 7.4. Benchmark datasets and models

This study uses the CORD-19 dataset. We acknowledge that CORD-19 data is being used in many recent works (Ngai et al. 2021; Bhatia et al. 2020), however, compared to the previous works, we use the latest release of CORD-19 as of December 2021. The other recent works use the releases of CORD-19 that are not so recent, so many topics, such as vaccination, long-COVID, post-COVID-19 symptoms and impacts may not be covered in those works. Our goal is to keep our research on the ongoing COVID-19 issues as well as to address the post-COVID-19 impacts.

NIH COVID-19 Portfolio[20], CORD-19 (Lu Wang et al. 2020) and LitCOVID (Chen, Allot, and Lu 2021) are among the initial efforts for COVID-19 datasets, and we find many other datasets since been, including, Covidex (E. Zhang et al. 2020) and others ,as listed in recent literature (Chen et al. 2021; Wang and Lo 2021). However, according to our preliminary empirical research, there is a significant overlap of articles in these datasets, which is understandable given that the COVID-19 articles are available in each of these datasets or scholarly repositories. CORD-19, for example, focuses on publications, WHO documents, and pre-prints. LitCOVID only contains articles that have been published in journals and don't include pre-prints, but there is a huge overlap between these two datasets.

We chose CORD-19 for its broad coverage of articles; additionally, the overall goal of the study is to propose the design of a search engine based on COVID-19, so this dataset is a good fit for our study. CORD-19, like LitCOVID, provides updates (latest releases) for the articles in package form (zip formats) and an API to filter the topics, so we worked to extract the text data from the files within those packages. In future, we would like to consider additional datasets, keeping in mind that duplicate or overlapping data should be avoided.

One future direction in this line of research is to prepare an aggregate dataset from other potential sources, such as publishers, such as Elsevier's Novel Coronavirus Information Center[21], Springer Nature's Coronavirus Research Highlights[22], or JAMA Network's COVID-19 Collection[23], which provide COVID-19 literature under temporary open access licenses through PMC's Public Health Emergency COVID-19 Initiative[24]. However, a significant challenge in generating a dataset from publisher websites is that full text may be unavailable in some cases or may only be available in the form of PDFs, which require extensive preprocessing to extract full text. Also, the open access status of many articles in these journals is unclear, which may result in unpleasant license revocations and sparsity of dataset and system failure in the future.

[20] https://icite.od.nih.gov/covid19/search/
[21] https://www.elsevier.com/connect/coronavirus-information-center
[22] https://www.springernature. com/gp/researchers/campaigns/coronavirus
[23] https://jamanetwork.com/journals/jama/ pages/coronavirus-alert
[24] https://www.ncbi.nlm.nih.gov/pmc/about/ covid-19/

We evaluate CO-SE using two benchmark datasets: COVID-QA and BioASQ, in addition to our CQuAD data to evaluate the performance of CO-SE. COVID-QA is not quite updated, but BioASQ is quite a recent benchmark dataset for COVID-19-related question-answering tasks, however, according to our initial assessment BioASQ coverage is limited. We understand that COVID-19 is a novel subject, and benchmark datasets in SQuAD format have been scarce until now, that's why our evaluation is limited to three datasets. We also attempt to prepare a SQuAD dataset for COVID-19 (CQuAD), which is currently quite small due to resource constraints (annotators, computational resources). In the future, we intend to continue this line of research by benchmarking this dataset for COVID-19 research and would like to invite researchers to join us in this endeavour. Additionally, we would like to explore additional datasets that may become available in near future.

There are some related works on COVID-19 question-answering and information retrieval systems in the state-of-the-art. CovidQA (Tang et al. 2020), COVID-QA system (Möller et al. 2020), COVIDASK (Lee et al. 2020) and listed in recent reviews (Chen et al. 2021; Wang and Lo 2021), however, many of these models focus on early and mid-pandemic. Our work focuses on long-COVID and other issues, such as the impacts of COVID-19 on different population groups, besides the early and mid-pandemic issues. In addition, our gold-standard dataset consists of articles on long-COVID, vaccines, equity and socio-economic topics related to COVID-19 and we have released the model weights of our student model trained on this dataset that can be used for building a similar reader component. We contribute to advancing the ongoing pandemic research in artificial intelligence to the next step.

### 7.5. Transfer learning, pre-training and fine-tuning

In this work, we fine-tune the DistilBERT with two research goals: (1) to prepare a student model that inherits the benefits of its predecessor models (e.g., BERT) and (2) to fine a Transformer model on our gold-standard dataset. We acknowledge that pre-training a model on a dataset (e.g., COVID-19 in this work) could give us a more domain-specific model, however, we are constrained by two things for pre-training: (1) pre-training requires a lot of computational resources, which we don't have in current setup; (2) our gold-standard dataset size is too small to pre-train a model from scratch, it has around 150 question-answering pairs.

Given the scenario that we have the resources, and if we enforce the model to pre-train on a smaller data, it may lead to underfitting or overfitting (Weiss, Khoshgoftaar, and Wang 2016). Two future directions, in this regard, are (1) to prepare a larger annotated data (maybe thousands of question-answering pairs) in SQuAD format; (2) given more computational resources (such as hardware, software, time availability), pre-train a Transformer model (BERT, GPT2, T5, BART or any alternative) on the annotated dataset. The power of these language models (such as BERT, GPT-2 and so) comes from training a large amount of data, so pre-training can be very useful in future.

*Conclusion:* we discuss the practical implications of CO-SE in this section, we also suggest some future directions for the researchers to extend this model. We hope that this resource will continue to bring together members of the computing community, biomedical specialists, and policymakers in the pursuit of effective treatments and cures for COVID-19 and to ending the pandemic and fighting any future pandemics.

## 8. Conclusion

In this paper, we propose the CO-SE system as a solution to the COVID-19 challenges, assisting researchers and clinical workers in obtaining scientific information in the form of search engine results. The architecture gets its data from a wide variety of COVID-19 topics and also focuses on long-COVID. The core of the architecture is a pipeline that is composed of a document store that stores scientific papers from the CORD-19; a retriever that retrieves documents from the document store in response to a question; and a reader that extracts the specific answer to each query from the documents returned by the retriever. Additionally, the returned responses are ranked and evaluated for exact word and semantic similarity. The experimental results demonstrate our model's superiority over state-of-the-art models. This is due to the unique design of our model and the large amount of data collected in this research. In the future, we would like to extend this work to other related tasks, such as question answering, summarization, translation, and clustering and consider critical appraisal methods for evaluating the data's credibility. We also plan to prepare a dashboard for the proposed system.

**Data Availability**

Data will be available upon request.

**Authors' Contributions**

The first author has contributed to this manuscript.

**Conflicts of Interest**

There is no conflict of interest associated with this article.

**Acknowledgments**

I would like to acknowledge that this research and manuscript is a part of my CIHR Health Systems Impact Fellowship.